# A Random Forest and Current Fault Texture Feature–Based Method for Current Sensor Fault Diagnosis in Three-Phase PWM VSR

Lei Kou[1,2], Xiao-dong Gong[1,2,3], Yi Zheng[1,2]*, Xiu-hui Ni[1,2,3],

Yang Li[4], 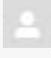Quan-de Yuan[5] and Ya-nan Dong[6]

- [1]Institute of Oceanographic Instrumentation, Qilu University of Technology (Shandong Academy of Sciences), Qingdao, China
- [2]Joint China-Ukrainian Scientific and Innovation Laboratory for Hydroacoustics, Qingdao, China
- [3]National Technical University of Ukraine "IgorSikorsky Kyiv Polytechnic Institute", Kyiv, Ukraine
- [4]School of Electrical Engineering, Northeast Electric Power University, Jilin, China
- [5]School of Computer Technology and Engineering, Changchun Institute of Technology, Changchun, China
- [6]Qingdao Branch, China Merchants Bank, Qingdao, China

Three-phase PWM voltage-source rectifier (VSR) systems have been widely used in various energy conversion systems, where current sensors are the key component for state monitoring and system control. The current sensor faults may bring hidden danger or damage to the whole system; therefore, this paper proposed a random forest (RF) and current fault texture feature–based method for current sensor fault diagnosis in three-phase PWM VSR systems. First, the three-phase alternating currents (ACs) of the three-phase PWM VSR are collected to extract the current fault texture features, and no additional hardware sensors are needed to avoid causing additional unstable factors. Then, the current fault texture features are adopted to train the random forest current sensor fault detection and diagnosis (CSFDD) classifier, which is a data-driven CSFDD classifier. Finally, the effectiveness of the proposed method is verified by simulation experiments. The result shows that the current sensor faults can be detected and located successfully and that it can effectively provide fault locations for maintenance personnel to keep the stable operation of the whole system.

# 1 Introduction

Three-phase PWM VSR systems have been incrementally applied in the field of renewable energy systems, uninterruptible power supplies (UPSs), motor drivers, and so forth and have become an indispensable energy conversion device (Yang et al., 2016; Li and Yang, 2017; Bueno and Pomilio, 2018; Zhang et al., 2021). However, most of the renewable energy equipment is built in the mountains or remote areas away from the coast as the probability of sensor faults is very high under harsh climates or conditions (Bidadfar et al., 2021). Once the fault occurs in the key sensors related to the control system, it may cause the whole system to crash or run inefficiently (Peng et al., 2018). Therefore, the sensor fault diagnosis for three-phase PWM VSR systems is of great significance to ensure the reliability of the whole system.

According to the fault degree, the sensor fault types can be divided into hard faults and soft faults (Huang and Tan, 2008; Li et al., 2011; Darvishi et al., 2021). Hard faults are usually caused by sensor components' damage, or electrical system short circuit or open circuit, and the measured value will change greatly. Soft faults generally refer to the aging of sensor components. The harm of soft faults is not as great as that of hard faults, but long-term operation will reduce the efficiency and accelerate the aging of systems (Fravolini et al., 2019). For example, if the fault output signals of the current sensors are used as the input signals of the control system, it will affect the closed-loop feedback control, reduce the control performance, accelerate the aging of other equipment, and even lead to safety accidents. If the early soft fault features are monitored, the potential hazards can be found in time, the maintenance can be carried out in time, other health equipment can be protected, and the stability of the whole system can be ensured. Therefore, an accurate and effective fault diagnosis method for sensor faults is especially essential to enhance the reliability of the system, and it would be better if the fault locations can be detected.

Sensor fault detection and diagnosis (SFDD) methods can be broadly divided into data-driven and model-based methods (Reppa et al., 2015; Lee et al., 2021). The model-based methods are usually easy to integrate into control systems, but they also need to set complex thresholds, and the model-based methods are more difficult to apply in different fields, especially for a nonlinear system since the fault models are more difficult to establish (Kou et al., 2020; Wang et al., 2020). The data-driven methods can only use the historical data to establish a black-box model and use the mature black-box classifier to realize fault diagnosis and location, without the requirement of understanding the fault mathematical model about the sensor systems (Li F. et al., 2021; Shi et al., 2008). Therefore, the data-driven methods do not rely on mathematical models and have attracted attention of many scholars (Ojo et al., 2021). Lee et al. (2021) proposed a convolutional neural network (CNN)–based FDD method for battery energy storage systems to detect and classify false battery sensor data. Ojo et al. (2021) proposed a long short-term memory recurrent neural network (LSTM-RNN)–based thermal fault diagnosis method for lithium-ion batteries, which is an easy-to-implement way and does not need to pay attention to the complex mathematical modeling and parameters of battery physics. Chen et al. (2021a) proposed a hypergrid and statistical analysis–based FDD method, which can identify the sensor faults in wireless sensor networks. Jana et al. (2021) developed a distributed SFDD framework, in which a fuzzy deep neural network (FDNN) was adopted to detect and identify the sensor faults. Hajer and Okba (2020) proposed an interval-valued data-driven method, which was adopted to detect and locate the sensor faults in chemical industrial fields. Gao et al. (2020) proposed a CNN-based FDD method for micro-electromechanical system (MEMS) inertial sensors, in which the time-domain features of temperature-related sensor faults were adopted to train the data-driven FDD classifier. Haldimann et al. (2020) proposed a disentangled RNN and residual analysis–based SFDD method and developed a novel procedure to identify the fault sensors. Chen et al. (2021b) proposed an NN-based fault estimation method, which can obtain the accurate estimation of sensor faults. Li L. et al. (2021) proposed an LSTM-based CSFDD method, which can learn data features automatically and predict the acceleration responses from the measured data. According to the previous discussion, data-driven methods have been widely used for sensor fault diagnosis in various fields. Data-driven methods can effectively set up the nonlinear model between input features and fault modes, and the historical fault data under both normal and fault modes can be obtained from the simulation tools. However, the research on fault data extraction is very important, which is helpful to improve the fault diagnosis accuracy of the whole diagnosis system.

Many scholars have made good achievements in sensor fault diagnosis, which have given us many experiences. Although the artificial neural network (ANN) is a popular supervised learning algorithm, it is vulnerable to causing over-fitting and affecting the generalization ability and the diagnosis results. Therefore, the random forest (RF) algorithm is adopted to train the CSFDD classifier, which is not easy to fall into over-fitting due to the introduction of two randomness (random samples, random features) (Roy et al., 2020; Fezai et al., 2021). Meanwhile, the current fault texture features are proposed to train the RF CSFDD classifier, which can improve the feature diversity and the diagnosis accuracy. Hard faults can cause severe damage, and it usually takes a long time to change from soft faults to hard faults. Therefore, if soft faults can be detected and diagnosed in time, the equipment can be maintained in time and hard faults can be avoided.

As shown in Figure 1, the proposed method can be applied to the sensor fault diagnosis of power-electronics energy conversion systems in the photovoltaic system, the wind generation system, and so forth. Therefore, this paper mainly takes the soft faults and hard faults of current sensors in the three-phase PWM

VSR as examples to verify the proposed CSFDD method. The detailed contributions of this study are summarized as follows:

(1) Only three-phase alternating currents (ACs) are selected as the input signals, additional sensors will not be introduced, and external interference can be avoided.
(2) It presents a current fault texture feature–based method, which can retain important fault features.
(3) A mature data-driven CSFDD classifier is trained by the RF algorithm and extracted current fault texture features, which can realize fault diagnosis and location for current sensors.

Figure 1

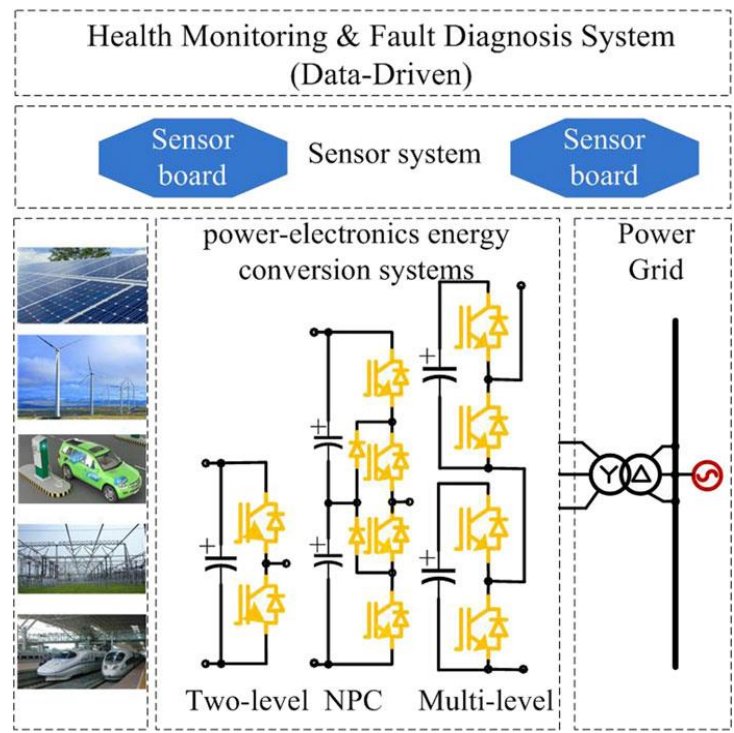

FIGURE 1. Sensor fault diagnosis and health monitoring system.

The remainder of this paper is organized as follows. Section 2 describes the current sensor fault data and fault features. In Section 3, the current fault texture feature–based method is proposed and the mature RF CSFDD classifier is trained and compared with the classifier trained by original samples. In Section 4, the proposed method is verified by simulation experiments. Conclusions are drawn in Section 5.

# 2 Current Sensor Data of Three-Phase PWM VSR

## 2.1 Structure of Three-Phase PWM VSR

The main circuit topology of the three-phase PWM VSR is depicted in Figure 2. As shown in Figure 3, the control strategy of the three-phase PWM VSR in this article is proportional resonance (PR), and it can be referred from Rocha et al. (2018).

Figure 2

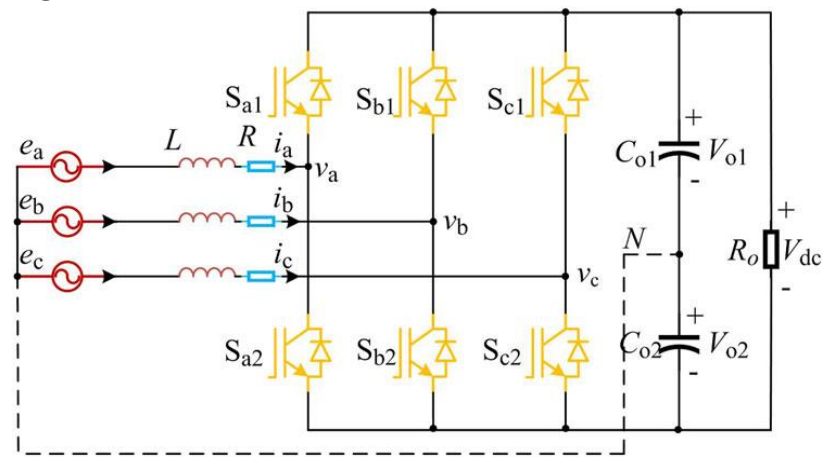

FIGURE 2. Main circuit topology of the three-phase PWM VSR.

Figure 3

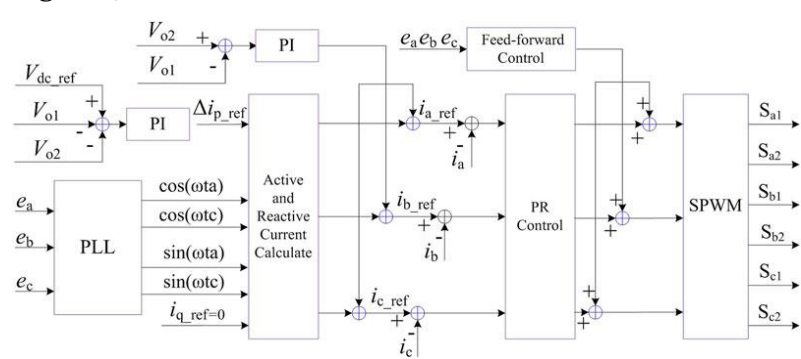

FIGURE 3. Basic control block diagram.

Under ideal conditions, [Math Processing Error]Sk stand for the switching functions, whose equation can be expressed as where [Math Processing Error]Sk1 and [Math Processing Error]Sk2 stand for the upper and lower IGBTs' switching state of each phase, respectively.

The expression of [Math Processing Error]vk can be described as where the relational expressions of [Math Processing Error]Vo2 and the output voltage [Math Processing Error]Vdc can be expressed as

The dynamic model of the three-phase PWM rectifier can be expressed as follows:

According to Eqs 1–4, the mathematical models in the a–b–c frame for the three-phase PWM rectifier can be expressed as

## 2.2 Current Sensor Fault Data

The three-phase currents of the three-phase PWM VSR are collected. Figure 4 shows the experimental platform of the three-phase PWM VSR, where the input phase voltage is 40 V, the grid voltage frequency is 50 Hz, the load is 16 [Math Processing Error]Ω, the sampling frequency is 25.6 kHz, and the output voltage is 100 V. Figure 5 shows the three-phase AC data under normal conditions.

Figure 4

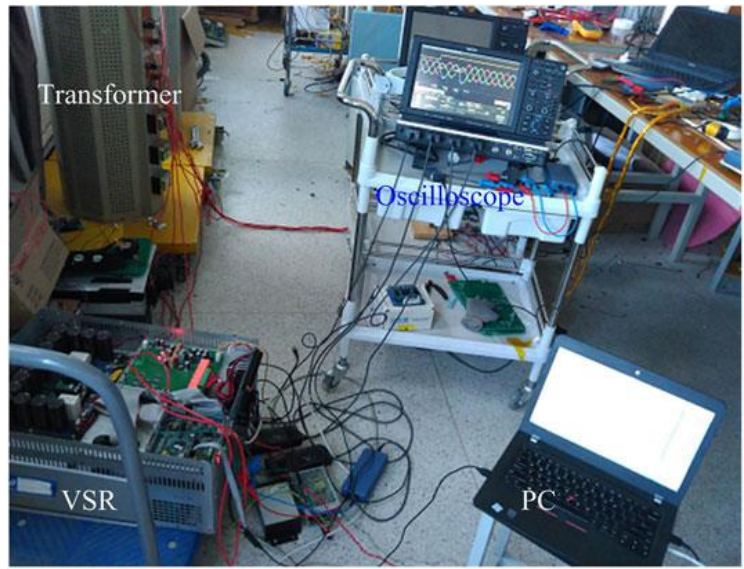

FIGURE 4. Experimental platform.

Figure 5

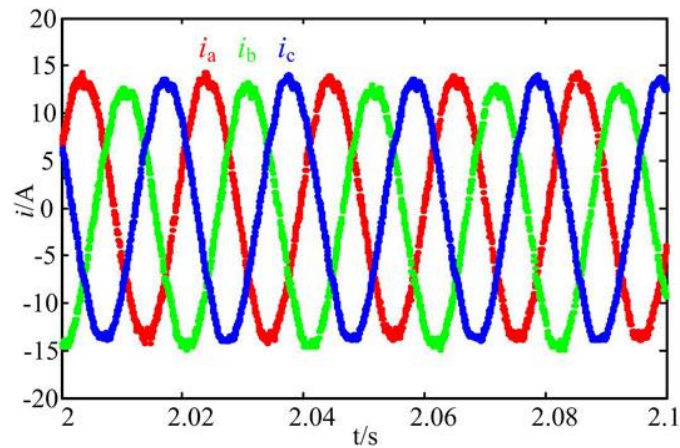

FIGURE 5. Three-phase current data under normal conditions.

Suppose the expression of normal three-phase currents is expressed as where [Math Processing Error]ia, [Math Processing Error]ib, and [Math Processing Error]ic are the three-phase currents, A is the amplitude of three-phase currents, and [Math Processing Error]ω=100π.

When the faults occur in the current sensors, the measured three-phase currents will be randomly increased by a number of different deviations. The measured three-phase currents can be expressed as

According to the research by Li et al. (2011) and Huang and Tan (2008), the soft faults and hard faults are simulated. Taking the soft fault and the hard fault of the A-phase current sensor as examples, the soft fault is simulated by reducing the measured A-phase current to 70% of the normal value (as shown in Figure 6) and the hard fault is simulated by reducing the measured value to 40% of the normal value (as shown in Figure 7). Meanwhile, the random noises are also superimposed. Figure 8 shows the three-phase current data with multiple current sensor faults, Figure 8A shows the three-phase current data when soft faults occur in both A-phase and B-phase current sensors, and Figure 8B shows the three-phase current data when soft fault occurs in the A-phase current sensor and hard fault occurs in the B-phase current sensor. According to Figures 6–8, in some cases, when faults occur in multiple current sensors at the same time, the fault phase currents will have superposition effect of fault features. According to Figures 6, 7, soft faults have little harm in a short time, but they will bring great harm when these faults develop into hard faults. Therefore, the soft faults of current sensors should be monitored in time to avoid hard faults.

Figure 6

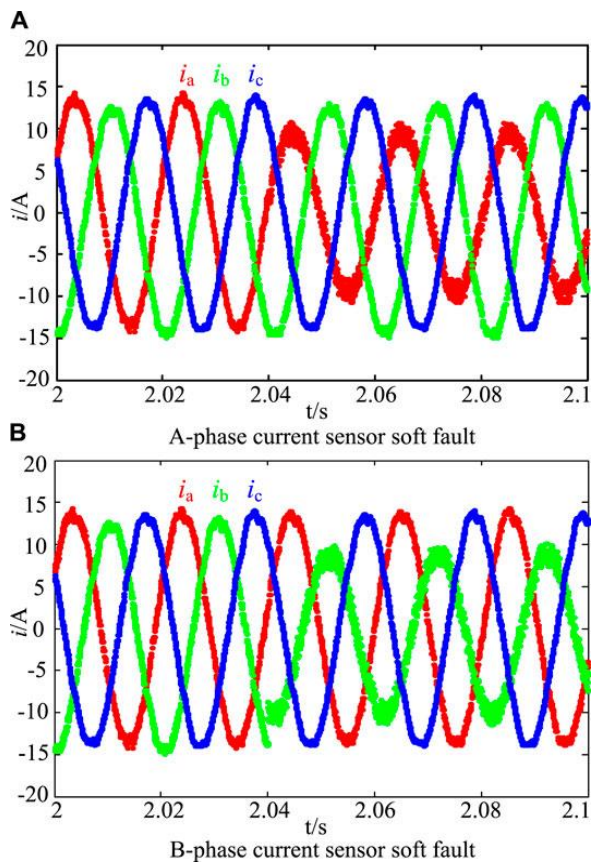

FIGURE 6. Three-phase current data with current sensor soft fault.

Figure 7

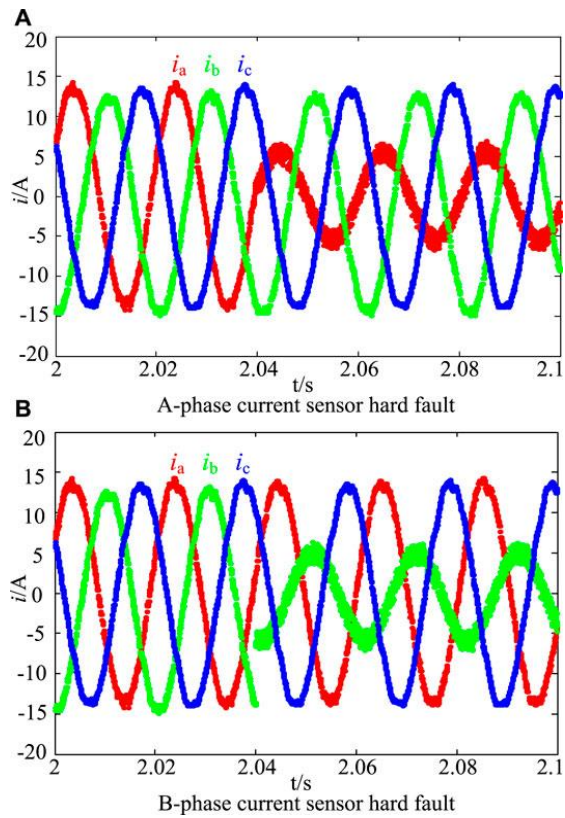

FIGURE 7. Three-phase current data with current sensor hard fault.

Figure 8

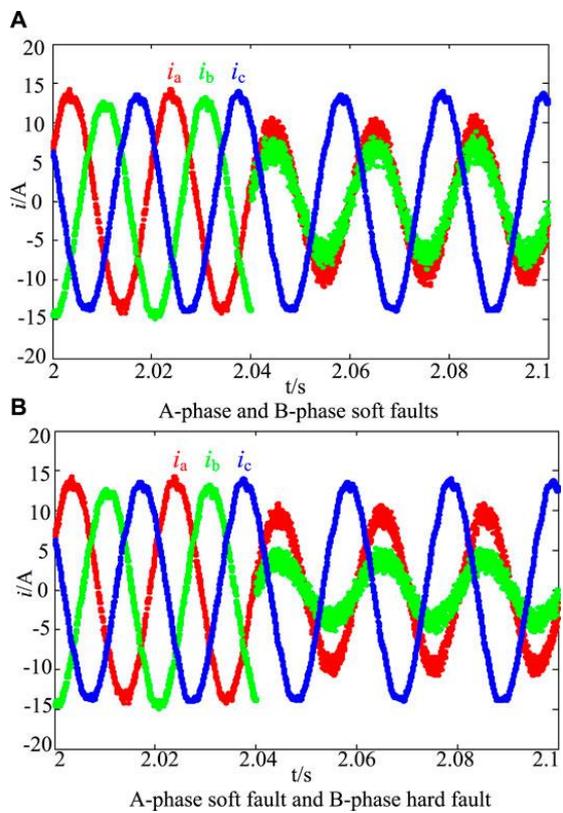

FIGURE 8. Three-phase current data with multiple current sensor faults.

# 3 CSFDD Classifier Based on RFs and Current Fault Texture Features

In this section, the current fault texture features are proposed, and then the fault diagnosis method based on RFs and current fault texture features is described. And meanwhile compared with the CSFDD classifier trained with original samples, the CSFDD classifier trained with current fault texture feature samples obtained a higher accuracy.

## 3.1 Current Fault Texture Features

Current fault texture features are important attributes of three-phase ACs, which can reduce the data dimension, retain most of the main features, and improve the ability of anti-interference. The main current fault texture features include the mean value, standard deviation, smoothness, skewness, kurtosis, and root mean square value. The current fault texture features retain most of the important features of three-phase ACs in a cycle.

## 3.2 Training and Evaluation of the CSFDD Classifier

The RF algorithm is a widely used ensemble learning algorithm, which can make joint decision through multiple weak classification models to improve the accuracy of decision (Gu et al., 2020). In the training process of RFs, samples and features are randomly selected in order to increase the difference between each weak model; the tree model can be more diversified by this way. The bagging method is usually used in the RF algorithm (Song et al., 2016). In the bagging method, the independence of different models is improved by using different independent training datasets. Through random sampling with the replacement method, the training sample set is extracted from the original sample set, and n training samples sets are extracted from the original sample set each time. In the training sets, some samples may have been repeatedly extracted and some samples may have never been selected. Using n training sets to train n models, n models are independent of each other, and then the integration model is obtained by voting.

where *[Math Processing Error]$\varepsilon k(X)=hk(X)–f(X)$* is the *[Math Processing Error]kth* error of the input samples *X*. And the average error of all weak models is

The direct average method is also adopted to integrate all weak classification models, and the expression of the integrated model is

The expected error of the integrated model is

According to the above derivation, the greater the difference between all weak models is, the better the performance of the integrated model will be. At the same time, the error will gradually decrease and tend to zero with the increase in the number of models.

The decision tree is a widely used classification model in the classification algorithm, where the classification and regression tree (CART) has the advantages of fast training speed, good stability, and supporting multiple segmentation of continuous data and feature data. Bagging algorithm plays a key role in the construction of random forest classifier, and its process is shown in Table 1. As shown in Figure 9, the CART is often used as the basic learning algorithm of RFs, whose structure is a binary classification tree.

Table 1

(1) Input: Sample training set $S=(X^1, Y^1), (X^2, Y^2), \ldots, (X^N, Y^N)$,
$X^N$ represents the feature attribute, $Y^N$ represents the sample label;
The total number of training is M, $\xi$ is the function of the weak classifier,
$S_{sub-j}$ is the selected sample training set;
(2) for $j = 1, 2, 3, \ldots, M$
(3) $h_j(X) = \xi(S, S_{sub-j})$;
(4) end
(5) Output: $H(X) = \text{argmax}\left( \sum_{j=1}^{M} \prod h_j(X) = Y \right)$

TABLE 1. Process of the bagging classification algorithm.

Figure 9

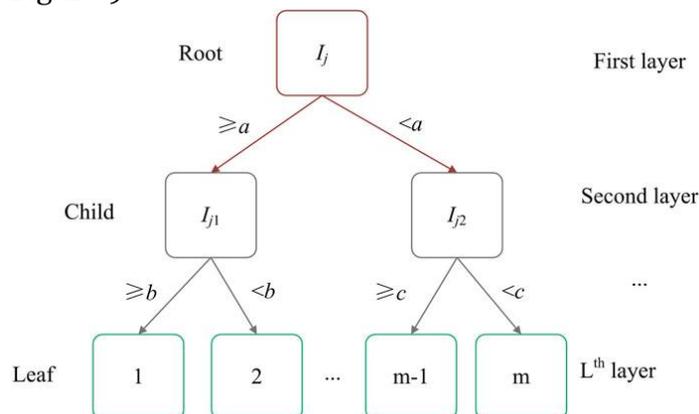

FIGURE 9. CART structure.

The purpose of the RF classifier is adopted to make a useful model with better classification performance, and it will formulate some set of rules when inputting a training dataset with features $X$ and labels $Y$ into the RF algorithm (as shown in Figure 10). What is most mysterious is that it can correctly predict the target class label of the sample in the testing datasets, which are given only the features without class labels. Figure 10 shows the original training flow chart of the RF algorithm with original features, and Figure 11 shows the training flow chart of the RF algorithm with current fault texture features. The biggest difference between them is that the latter introduces the current fault texture feature–based method to transform the original data features, where the current fault texture features are obtained after the transform of three-phase currents. The current fault texture feature–based method is adopted to perform feature extraction, and the data-driven method is used to train the CSFDD classifier with the extracted features.

Table 2

| Fault sensors | Class labels | | | | | |
|---|---|---|---|---|---|---|
| | S1 | S2 | S3 | H1 | H2 | H3 |
| Normal state | 0 | 0 | 0 | 0 | 0 | 0 |
| A-phase soft fault | 1 | 0 | 0 | 0 | 0 | 0 |
| A-phase hard fault | 0 | 0 | 0 | 1 | 0 | 0 |
| B-phase soft fault | 0 | 1 | 0 | 0 | 0 | 0 |
| B-phase hard fault | 0 | 0 | 0 | 0 | 1 | 0 |
| C-phase soft fault | 0 | 0 | 1 | 0 | 0 | 0 |
| C-phase hard fault | 0 | 0 | 0 | 0 | 0 | 1 |
| A–B-phase soft faults | 1 | 1 | 0 | 0 | 0 | 0 |
| A–B-phase hard faults | 0 | 0 | 0 | 1 | 1 | 0 |
| A-phase soft fault and B-phase hard fault | 1 | 0 | 0 | 0 | 1 | 0 |

TABLE 2. Fault sensors and class labels.

Figure 10

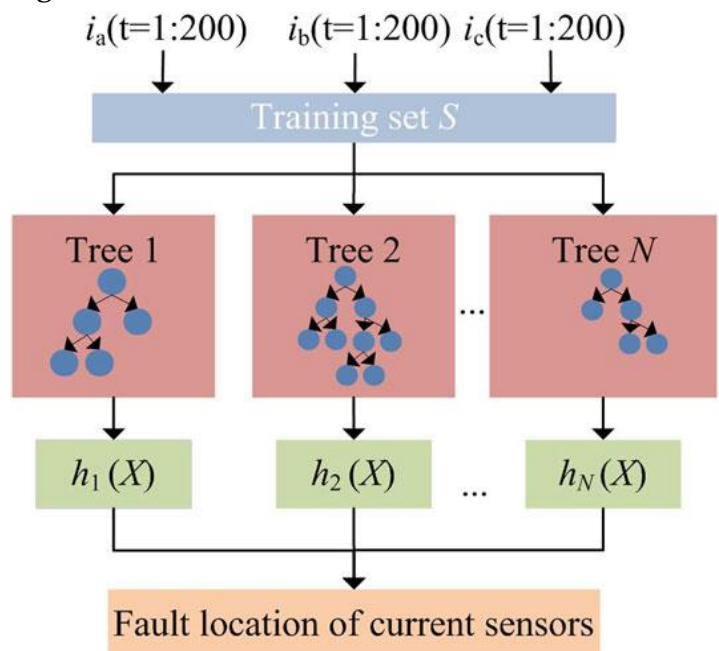

FIGURE 10. Original training flow chart.

Figure 11

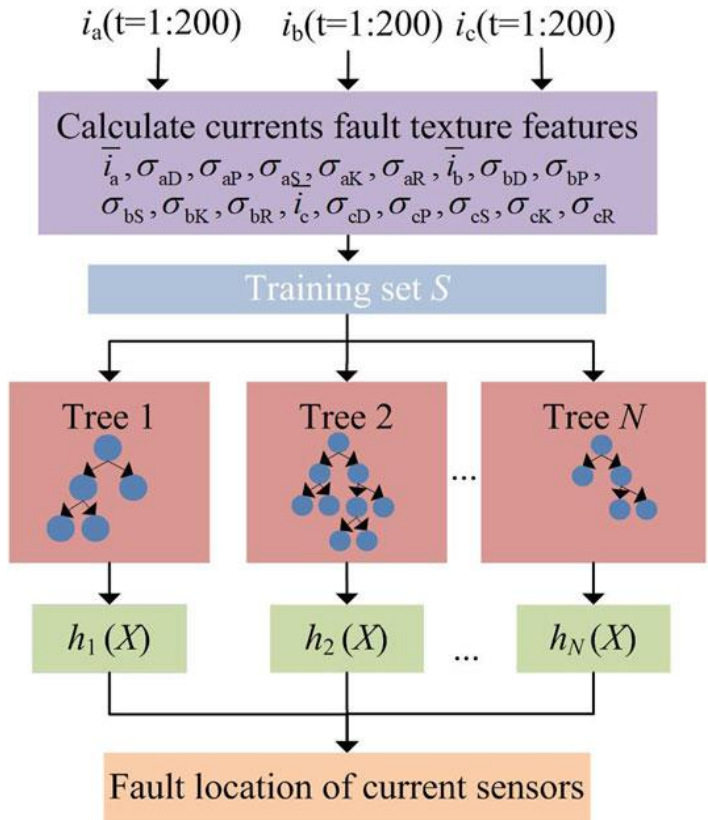

FIGURE 11. Training flow chart with current fault texture features.

In the training process of the CSFDD classifier, the training set is formed by randomly selecting 16,800 samples from 24,000 samples under different operating conditions, and the rest samples are used as the test samples. Here, the repeated cross test method is adopted to train the RF classifier. Figure 12A shows the trend of diagnosis results varying with trees when the CSFDD classifier was trained with original features, and when the number of RF decision trees is set as 305, the highest accuracy is 0.9636. Figure 12B shows the trend of diagnosis results varying with trees when the CSFDD classifier was trained with current fault texture features, and when the RF trees' number is set as 209, the highest accuracy is 0.9715.

Figure 12

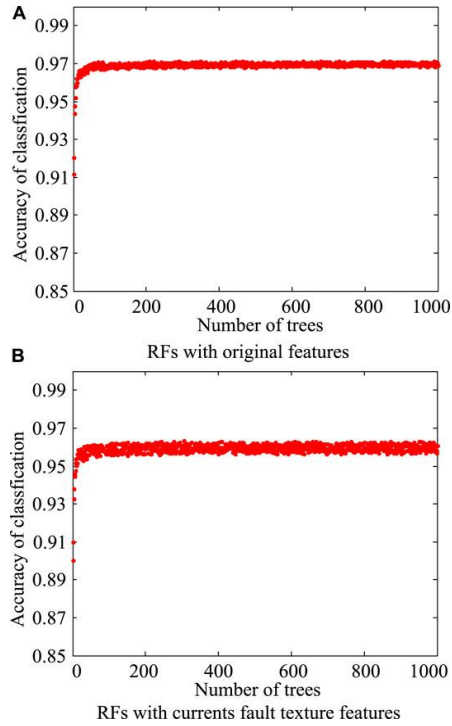

FIGURE 12. Influence of decision tree on performance of RFs.

Table 3 shows the diagnosis accuracy of RFs with original features, in which the CSFDD classifier was trained with the original fault current samples (as shown in Figure 10). Table 4 shows the diagnosis accuracy of the proposed method, in which the CSFDD classifier was trained with current fault texture features (as shown in Figure 11). According to Tables 3, 4, since the original fault current samples contain more noise, and the current fault texture features retain a large number of important features and remove noise, the accuracies of the proposed method are higher than those of RFs with original features. The proposed method offers high accuracy even with a training dataset of 70% of total samples compared with the test/validation dataset, which gives more than 97.15% average accuracy for different types of faults.

Table 3

| Fault sensors | Diagnosis accuracy | Misdiagnosis rate |
| --- | --- | --- |
| Normal state | 0.9671 | 0.0329 |
| A-phase soft fault | 0.9606 | 0.0394 |
| A-phase hard fault | 0.9652 | 0.0348 |
| B-phase soft fault | 0.9486 | 0.0514 |
| B-phase hard fault | 0.9673 | 0.0327 |
| C-phase soft fault | 0.9526 | 0.0474 |
| C-phase hard fault | 0.9611 | 0.0389 |
| A–B-phase soft faults | 0.9643 | 0.0357 |
| A–B-phase hard faults | 0.9606 | 0.0394 |
| A-phase soft fault and B-phase hard fault | 0.9666 | 0.0334 |

TABLE 3. Accuracy of RFs with original features.
Table 4

| Fault sensors | Diagnosis accuracy | Misdiagnosis rate |
|---|---|---|
| Normal state | 0.9745 | 0.0255 |
| A-phase soft fault | 0.9743 | 0.0257 |
| A-phase hard fault | 0.9650 | 0.0350 |
| B-phase soft fault | 0.9663 | 0.0337 |
| B-phase hard fault | 0.9752 | 0.0248 |
| C-phase soft fault | 0.9715 | 0.0285 |
| C-phase hard fault | 0.9725 | 0.0275 |
| A–B-phase soft faults | 0.9735 | 0.0265 |
| A–B-phase hard faults | 0.9675 | 0.0325 |
| A-phase soft fault and B-phase hard fault | 0.9730 | 0.0270 |

TABLE 4. Accuracy of RFs with current fault texture features.

# 4 Simulation Experiments

To verify the validity of the CSFDD classifier based on RFs and current fault texture features, the simulation experiments of soft and hard faults are carried out, and the soft fault and hard fault of the A-phase current sensor and multiple current sensor faults are studied as examples. Figure 13 shows the flow chart of sensor fault diagnosis, where the current samples are five cycles of fault samples, and it contains five sets of samples. When a set of fault samples are input each time, they will be processed by the current fault texture feature method and then input into the mature CSFDD classifier to get the final diagnosis results. The data of Figures 6A, 7A, 8 are input into the mature CSFDD classifier (samples of one cycle are input at a time, as shown in Figure 13), and then the location and type of the fault sensor can be diagnosed.

Figure 13

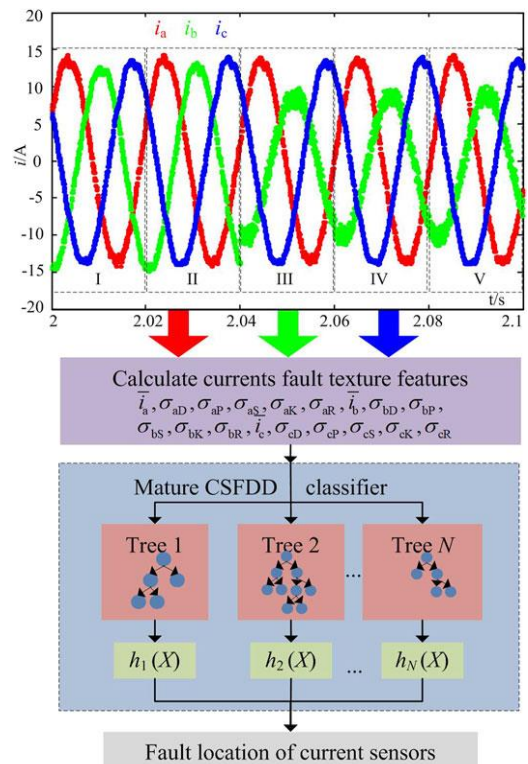

FIGURE 13. Flow chart of sensor fault diagnosis.

Figure 14 shows the diagnosis results of five cycles. Figure 14A shows the diagnosis results of Figure 6A; at the beginning of diagnosis, the results are normal and the sensors work normally. When the soft fault occurs in the A-phase current sensor, the diagnosis results turn to A-phase soft faults, and the diagnosis results are consistent with the actual results. Figure 14B shows the diagnosis results of Figure 7A; the early diagnosis results are normal, the diagnosis results turn to A-phase hard faults when the hard fault occurs in the A-phase current sensor, and the diagnosis results are consistent with the actual results.

Figure 14

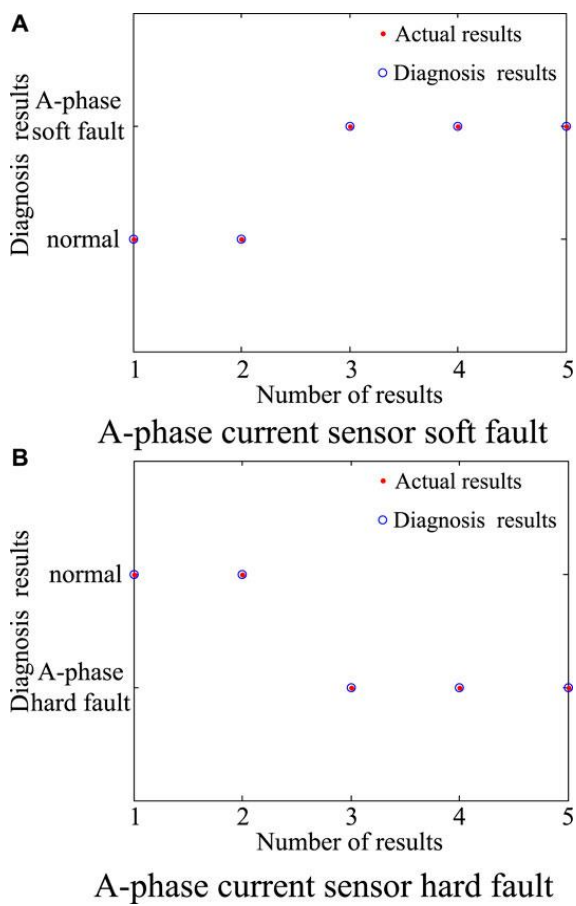

FIGURE 14. Fault diagnosis results with a single current sensor fault.

Figure 15 shows the fault diagnosis results with multiple current sensor faults. Figure 15A shows the diagnosis results of Figure 8A, and 15B shows the diagnosis results of Figure 8B. According to the diagnosis results, the proposed method can locate multiple faults and different types of faults.

Figure 15

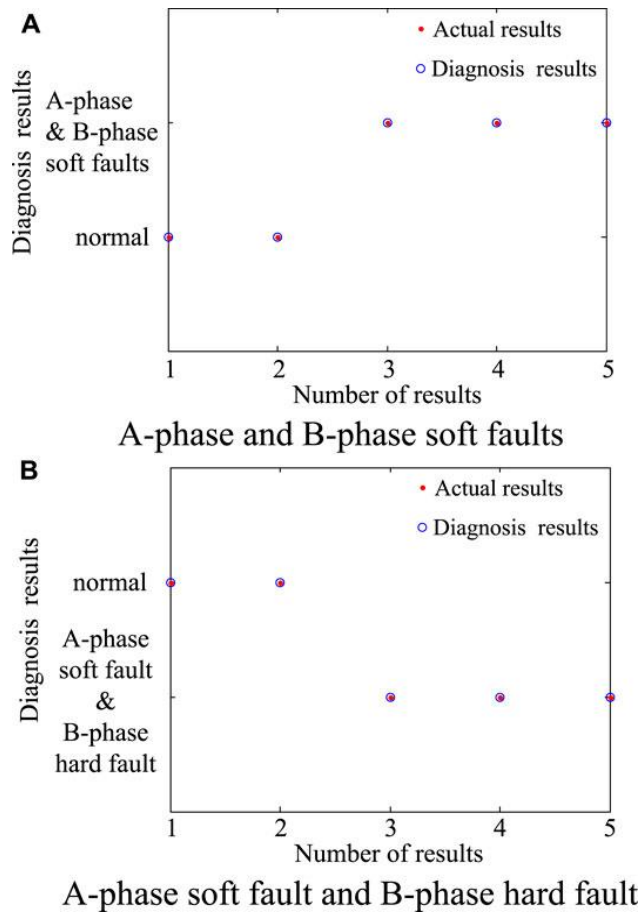

FIGURE 15. Fault diagnosis results with multiple current sensor faults.

According to the diagnosis results of Figures 14, 15, the current sensor faults can be effectively diagnosed and located by the proposed method.

# 5 Conclusion

This paper proposed an RF and current fault texture feature–based method for current sensor fault diagnosis in three-phase PWM VSR systems. The proposed method employed only three-phase ACs without other additional hardware. Firstly, the current fault texture feature–based method is adopted for feature extraction, which can retain a large number of important features and remove noise. After that, the current fault texture features are used to train the CSFDD fault classifier by the RF algorithm, and the accuracy of the proposed method is higher than those of RFs with original features. Finally, the results of simulation experiments show that the current sensor faults can be effectively diagnosed and located by the proposed method, and the experimental results prove the effective and robust performance of the proposed method.